\documentclass{aa}  

\usepackage{graphicx}
\usepackage{txfonts}
\usepackage{mathptmx}
\usepackage{amsmath}
\usepackage{amssymb}
\usepackage[T1]{fontenc}
\usepackage{ae,aecompl}
\usepackage{soul} 
\usepackage{sidecap}
\usepackage{dirtytalk}  
\usepackage{color}

\newcommand{\Msun}{M$_{\odot}$}
\newcommand{\Lsun}{L$_{\odot}$}

\newcommand{\Myr}{M$_{\odot}$\,yr$^{-1}$}

\usepackage{adjustbox}
\usepackage[graphicx]{realboxes}
\usepackage{rotating}
\newcommand{\Gabs}{G$_{\rm{abs}}$}
\newcommand{\G}{{\it Gaia}}

\begin{document} 

   \title{The \G\ Catalogue of Galactic AGB Stars}

   \subtitle{I. OH/IR stars}

   \author{B. L\'opez Mart\'{\i}\inst{1,2}
          \and
          F. M. Jim\'enez-Esteban\inst{3}
          \and
          D. Engels\inst{4}
          \and
          P. Garc\'{i}a-Lario\inst{5}
          }

   \institute{Universidad San Pablo CEU, E-28925 Alcorcón, Madrid, Spain\\
              \email{belen.lopezmarti@ceu.es}
         \and
         Spanish Virtual Observatory (SVO)
         \and
             Centro de Astrobiolog\'{\i}a (CAB), CSIC-INTA, Camino Bajo del Castillo s/n, E-28692, Villanueva de la Ca\~{n}ada, Madrid, Spain.
         \and
         Hamburger Sternwarte, Universit\"at Hamburg, Gojenbergsweg 112, D-21029 Hamburg, Germany
         \and
         European Space Agency (ESA), European Space Astronomy Centre (ESAC), Camino Bajo del Castillo, s/n, Urbanización Villafranca del Castillo, E-28692, Villanueva de la Cañada, Madrid, Spain
             }

   \date{Received month day, 2024; accepted month day, year}
 
  \abstract
   {The \G\ mission discovered several hundred thousand long-period variables and measured parallaxes for many of them. These stars will allow us to study populations of variable stars in the Milky Way, including asymptotic giant branch (AGB) stars.}
   {This paper describes the identification of \G\ counterparts of a sample of oxygen-rich AGB stars with OH maser emission as a first step towards the compilation of a general \G\ Catalogue of Galactic AGB stars. With this catalogue, tests of evolutionary models for the AGB star population in the solar neighbourhood become feasible.}
   {We cross-matched AGB star candidates showing OH maser emission with the \G\ DR3 release using a cross-match with AllWISE and 2MASS as intermediate steps to avoid ambiguities. With the help of the Virtual Observatory, we retrieved photometric data from the near-ultraviolet to the far-infrared and built spectral energy distributions (SEDs) of the sources. The SEDs were fitted with theoretical models. The fit results, together with information from the literature,
   allowed us to clean the sample from non-AGB stars. For the AGB stars, bolometric fluxes were obtained. Distances based on \G\ parallaxes were used to derive the stellar luminosities.}
   {We identified unique \G\ counterparts for 1487 OH masers. Of these, 1172 had an unambiguous classification as AGB stars. These sources make up the \G\ OH/IR star sample. Parallaxes with relative errors $<$20\% and astrometric excess noise $<$1.5 mas were available for 222 OH/IR stars.}
   {The study of the AGB population in the solar neighbourhood is limited by the obscuration by circumstellar dust, as \G\ DR3 only provides parallaxes for a few of our candidates. The location of the OH/IR stars matches that of LPV discovered by \G\ in the (BP$-$RP; G$_{abs}$) diagram, but the OH/IR star sample is biased towards redder colours (BP$-$RP$>$4) mag and larger amplitudes ($>$1 mag in the G-band), which are typical for periodic large-amplitude Mira variables.}


   \maketitle


\section{Introduction}

The third data release (DR3) from the  \G\ mission includes astrometry and broad-band photometry for a total of 1.8 billion objects based on 34 months of satellite operations \citep{GaiaDR3}. A subset of $\sim$1.7 million stars have been identified as long-period variable (LPV) star candidates with amplitudes $>$0.1 mag in the \textit{Gaia} photometric G-band \citep{Lebzelter23}. This subset includes stars on the asymptotic giant branch (AGB), which are in general variable stars and are therefore a major component of the LPV population. The measurement of parallaxes by \G\ holds the promise that sizeable samples of AGB stars with reliable distances will become available in the Milky Way and especially in the solar neighbourhood. They can be used for an observational validation of AGB evolutionary models. However, the selection of AGB stars from LPV catalogues needs to remove stars of different nature, for example young stellar objects (YSOs), which can mimic the variability behaviour of AGB LPV stars \citep{Lucas24}. Validation samples of well-curated AGB stars are therefore valuable to test the selection criteria for AGB stars in LPV catalogues.

The majority of low- and intermediate-mass stars ($\sim 0.8 - 8$ \Msun) evolve through the AGB phase, where they experience high mass-loss rates ($\ga10^{-7}$\Myr). When the mass-loss rate surpasses 10$^{-6}$\,\Myr, the dust shell eventually becomes opaque to visible light \citep{Habing96}, hiding stars in the advanced AGB phase from detection by \G. The dust of the circumstellar envelope (CSE) absorbs the radiation of the central star and thermally re-emits it mainly in the infrared (IR). At the same time, the gas of the CSE can be the source of strong maser emissions, for example by the OH molecule at 1612, 1665, and 1667 MHz.
In this paper, we use the term `OH/IR stars' to refer to all AGB stars with OH maser emission. This includes the classical dust-enshrouded OH/IR stars as well as optically bright LPVs, for example Mira variables.

According to evolutionary models, at the end of AGB evolution a different chemical composition is expected for the stellar surface and in succession for the CSE, depending on their progenitor mass. At solar metallicities, the stars end carbon rich (C/O$>$1) for progenitor masses 1.5\,\Msun\,$\la$\,M\,$\la$\,3\,\Msun, while below and above this mass range, the stars terminate the AGB evolution with an oxygen-rich chemistry \citep{DiCriscienzo16, Ventura18}. Depending on the metallicity, these boundaries can vary by several tenths of a solar mass \citep{Tosi22}.

Although these models are commonly accepted and applied to explain the AGB population in the Magellanic Clouds \citep{Dell'Agli15b, Dell'Agli15a, Pastorelli19, Pastorelli20}, their application to Galactic AGB stars need to be validated, as the Milky Way has a higher metallicity. This validation is pending because the lack of reliable distances makes determinations of stellar luminosities and the subsequent matches of individual stars with evolutionary tracks uncertain. For a small sample of optically obscured oxygen-rich stars close to the end of AGB evolution (`extreme OH/IR stars') that are assumed to be located in the Galactic bulge, \cite{Jimenez-Esteban15} found stellar luminosities within the mass range in which the stars should already have been converted into carbon stars. In contrast to this result, \cite{Blommaert18} questioned the use of a common distance to these OH/IR stars and argued all OH/IR stars in the bulge have low-mass stars ($\sim$1.5 M$_\odot$) as progenitors. A local AGB sample with reliable \G\ parallaxes may allow us to test the models anew.

In addition to distances, the determination of reliable stellar luminosities in the Galaxy requires well-covered spectral energy distributions (SEDs) to avoid imprecise bolometric corrections. The SEDs  can today be built with unprecedented coverage from a wealth of photometric surveys that are easily accessible through the Virtual Observatory (VO). These catalogues cover wavelength ranges from the UV to the far-IR. Based on theoretical models of SEDs representative of the AGB evolutionary phase, gaps in the observational coverage of the SEDs can be filled, and bolometric corrections can be avoided completely. 

This work is the first step to create a \G\ Catalogue of Galactic AGB stars with reliable distances from \G-DR3, focusing on O-rich AGB stars identified by their OH maser emission.  In forthcoming works, we will add a sample of C-rich AGB stars and a sample of O-rich AGB stars without detections of OH maser emission. 

The paper is structured as follows. In Sect.\,\ref{sec:sample} we describe the identification of OH maser emitting AGB stars in \G-DR3. In Sect.\,\ref{sec:SEDs} we construct their SEDs using VO tools, obtain their bolometric fluxes from SED fitting, and determine stellar luminosities when trustworthy \G\ parallaxes are available. The final \G\ OH/IR star sample is defined in Sect.\,\ref{sec:cat}. This is followed by the description of the properties of the sample in Sect.\,\ref{sec:properties}. Finally, we present our conclusions in Sect.\,\ref{sec:conclusions}.

\section{Identifying OH maser sources in \G\ DR3}
\label{sec:sample}

As starting point for the construction of the \G\ OH/IR star sample, we used the database of circumstellar OH masers\footnote{\footnotesize \url{https://www.hs.uni-hamburg.de/maserdb}} by \cite{Engels15b} in its version 2.5 \citep{Engels24}. This database (hereafter, the OH maser database) is a compilation of masers that were predominantly detected at 1612 MHz until 2022, and it is restricted to stars in the Milky Way. Overall, the database contains 16,879 individual observations with detections of 2905 different objects. This number was decreased to 2857 sources after we removed some entries that were erroneous or duplicated. We call this sample of sources from the OH maser database `the OH maser sample'. 

The majority of sources in this sample are AGB stars. Based on the definition given in the Introduction, they are considered OH/IR stars in this paper.

The source coordinates in the OH maser database were obtained from many different bibliographic resources, and as a result, their accuracy largely varies, from milli-arcseconds to several arcminutes in the worst cases. Moreover, a direct cross-match of a catalogue of OH masers with an optical catalogue such as \G\ might yield too many spurious counterparts because counterparts of the OH masers are often strongly obscured optically. Therefore, matches with field stars, especially in regions with a high source density such as the Galactic plane, cannot be excluded. For these reasons, our cross-matching procedure followed intermediate steps to improve the source coordinates with the help of infrared catalogues and to ensure that the OH maser counterparts were detectable at shorter wavelengths. This procedure is explained below.

\subsection{Cross-match with infrared catalogues}
\label{sec:ir-xm} 

 Because of the substantial mass-loss rates, AGB stars are strong infrared sources and are generally detected by surveys such as the Infrared Astronomical Satellite (IRAS), the Two-micron All Sky Survey (2MASS), and/or the Wide-field Infrared Survey Explorer (WISE). This allows us to  improve the coordinates of the OH maser database. In addition, a steep rise in the spectral energy distribution in the near-infrared  (NIR) indicates that a source is deeply obscured by circumstellar dust and is therefore expected to have a faint or even undetectable optical (e.g. \G) counterpart. Furthermore, the infrared photometry might help us to distinguish between correct and spurious cross-matches, and to some extent, it might also be useful to identify the different types of objects included in the OH maser database. 

The catalogues we used in this first step of the cross-match procedure were the AllWISE Source catalogue \citep[3.4 -- 22~$\mu$m;][]{AllWISE} and the 2MASS Point Source catalogue \citep[1.2 -- 2.2~$\mu$m;][]{2MASS-PSC}. Both catalogues have a similar astrometric accuracy: In AllWISE, positions for non-saturated sources with a high signal-to-noise ratio (S/N; 8 < W1 < 12~mag) are accurate to about 50 mas. In the case of 2MASS, the source positions are accurate to 70-80 mas in the range 9 < $K_s$ < 14~mag and to 120 mas for brighter sources. The astrometric accuracy of fainter sources decreases monotonically with brightness. However, 2MASS has a better spatial resolution than AllWISE (1{\arcsec}/pixel versus 1.375{\arcsec}/pixel), and it may therefore resolve sources that are blended in the latter catalogue. This allowed us to improve the identification of the OH maser counterpart in the infrared in case of confusion in the AllWISE data or when the mid-infrared source was very bright, which sometimes made it difficult to correctly place the photocentre of the star on the WISE images.  The higher resolution of 2MASS also helped us to identify the most likely counterparts in the \G\ DR3 data, whose spatial resolution is even better.

In crowded areas or in areas with extended diffuse infrared emission, we were unable to reliably identify the counterparts in one or both of the infrared catalogues. This issue was especially severe towards the Galactic centre. Therefore, OH masers within 2 degrees of the Galactic centre were removed from the cross-match procedure. We removed 279 sources from the OH maser sample for this reason, and the cross-match was then made for the remaining 2578 sources.

We used a radius of 15{\arcsec} for the cross-match between the OH maser database and AllWISE, and we used a radius of of 3{\arcsec} between AllWISE and 2MASS. All matches were visually inspected to assess their reliability. In some cases, an incorrect counterpart had been assigned because it happened to be the one nearest to the input coordinates. Sometimes, the coordinates from the OH maser database were so inaccurate that the actual IR counterpart lay well beyond the search radius. If an object was found in the visual inspection that seemed to be the correct counterpart of the OH maser source, then this AllWISE or 2MASS counterpart was assigned to the source. 

Because of the shorter wavelengths, it was expected that 2MASS contained fewer counterparts to the OH maser catalogue than AllWISE: Some sources would be deeply obscured and not detectable in this range. On the other hand, all 2MASS counterparts of OH maser sources were expected to have counterparts in the AllWISE Source Catalog. However, in practice, we found a number of sources without an AllWISE counterpart that had a counterpart in 2MASS; these sources are visible in the WISE images, but lack an entry in the catalogue, probably due to issues in the source extraction (they are mostly highly saturated sources). 

Unique IR counterparts were identified for 2427 OH masers, but for 151 detected masers listed in the OH maser database, no counterpart was found or it was not possible to assign a reliable counterpart in either of the infrared catalogues. The reasons were manifold. Part of the OH maser sources may have no (unambiguous) infrared counterpart because the coordinates are grossly inaccurate ($>1\arcmin$) in the original literature or because the maser was misclassified and has an interstellar origin. In other cases, the identification of the correct IR counterpart was difficult due to crowding or blending with nearby sources or surrounding diffuse infrared emission.

\begin{table}
\caption[]{\label{tab:XM-results} Results of the cross-match of detected OH masers in the database of circumstellar OH masers \citep{Engels24} with the AllWISE, 2MASS, and \G\ DR3 catalogues.} 
\begin{center}
\footnotesize
\begin{tabular}{lrr}
\hline\noalign{\smallskip}
Sample &  N & Total \\
\noalign{\smallskip} \hline\noalign{\smallskip}
OH maser sample                  &      & 2857 \\
\;\; Galactic Center sources & 279  &      \\
Cross-match input                  &      & 2578 \\
\;\; \G\ DR3 identifications     & 1487 &      \\
\;\; IR identifications only       &  940 &      \\
\;\; No or non-unique IR identifications &  151 &      \\
\noalign{\smallskip} \hline
\end{tabular}
\end{center}
\end{table}

\subsection{Cross-match with \G\ DR3}
\label{sec:Gaia-xm} 

The next step was to cross-match the sample of OH maser sources with unique IR counterparts with the \G\ DR3 catalogue \citep{GaiaDR3} using the 2MASS coordinates as input. When the source had no 2MASS counterpart, the AllWISE coordinates were used. The search radius in \G\ DR3 was 1.5{\arcsec}. 

As with the infrared cross-matches, we assessed the reliability of the identified counterparts by visual inspection. Because \G\ does not provide any images, we used the images from the Digital Sky Survey 2 \citep[DSS-2;][]{dss} and the Pan-STARRS survey \citep{Chambers17} for this assessment. They cover a similar brightness range as \G , but the spatial resolution is far lower. 
We unexpectedly found a few sources with AllWISE counterparts that had a counterpart in \G\ DR3 even though they lacked a 2MASS counterpart. As for the sources with 2MASS counterparts but no AllWISE counterparts, the near-infrared source in these cases was visible in the 2MASS images, but was not included in the point-source catalogue.

When there were several possible counterparts, we checked the \G\ colours and favoured the counterpart with the reddest colours or, if the colours of the possible counterparts were similar, the closest source to the input position. In some cases, the reliability of the possible counterparts could not be assessed, and the source remained without a \G\ identification.

The cross-match with the \G\ DR3 catalogue resulted in 1487 OH maser sources with unique \G\ counterparts and 940 sources with no (unique) optical identification.

\subsection{Parallaxes and distances}
\label{sec:dist} 

Although 88\% of the OH maser sources with unique \G\ counterparts have parallax measurements, these are not reliable in many cases, either because they are negative and thus spurious (21\% of the sample), or because the astrometric quality is insufficient. 

Recent studies of \G\ distances for AGB stars have been made: \cite{Andriantsaralaza22} analysed the uncertainties of the Gaia parallaxes on these stars by comparing them with parallaxes measured with maser observations with very long baseline interferometry (VLBI). More recently, \cite{Bhattachary24} presented a novel method for estimating statistical distances to AGB stars using the SED shape and a brightness estimate. They then compared the obtained distances with the \G\ distances. Both studies came to similar conclusions: It is difficult for \G\ to provide reliable parallaxes for sources with relative parallax errors higher than 20\%, and the \G\ distances for distances beyond 2\,kpc is underestimated. 

Based on these results, we defined parallaxes as “good” when the relative parallax error was lower than 20\% and the astrometric excess noise was lower than 1.5\,mas. This latter parameter measures the disagreement, expressed as an angle, between the observations of a source and the best-fitting standard astrometric model \citep{Lindegren21}. Following these criteria, the parallaxes of only 277 sources (about 19\% of the sample) were regarded as “good”.

For all sources with a good parallax, we extracted the geometric distances from the catalogue of \citet{Bailer-Jones21}. These distances, which are calculated from \G\ DR3 data using a probabilistic approach, are expected to be more reliable than a simple inversion of the parallax. We did not consider the photogeometric distances that these authors also provided because they were computed using colour information based on a galaxy model, and it is highly unlikely that this prior is applicable to our extremely red sources.

\subsection{Overall cross-match results}
\label{sec:pcat} 

Based on the cross-match with the AllWISE, 2MASS, and \G\ DR3 catalogues, we identified 1487 sources from the OH maser sample with unique \G\ DR3 counterparts (hereafter, the \G\ sample), 277 of which (19\% of the sample) had distance estimates from \citet{Bailer-Jones21}. 

The majority of sources in the \G\ sample (1424, or 96\%) have counterparts in the near-infrared (2MASS) and the mid-infrared (AllWISE). Eight sources only have an AllWISE counterpart, and 55 sources (4\% of the sample) only have a 2MASS counterpart. 

In addition, 940 sources from the OH maser database without \G\ counterparts (about 33\% of the OH maser sample) had detections in AllWISE and/or 2MASS. Of these, 729 sources (78\% of this infrared sample) had counterparts in both catalogues, 178 (19\%) were included only in AllWISE, and 33 (about 4\%) were included only in 2MASS. The remaining 151 sources in the OH maser sample were not detected in either of these catalogues, had ambiguous counterparts, or were removed because they are located within 2 degrees of the Galactic centre. The sources without \G\ counterparts are not further considered here. Table \ref{tab:XM-results} summarises our cross-matching results. 

We recall that although the majority of sources in the OH maser database are expected to be oxygen-rich AGB stars, other types of sources such as post-AGB stars, red supergiants (RSG), or unrecognised YSOs are included as well. Therefore, additional steps were required to remove them from the \G\ sample. These steps included the construction and analysis of the spectral energy distributions (SEDs) of the sources. They are explained in the next sections.

\section{Spectral energy distributions}
\label{sec:SEDs}

The SEDs of the sources from the \G\ sample were built and analysed with the help of the Virtual Observatory SED Analyzer\footnote{\footnotesize \url{http://svo2.cab.inta-csic.es/theory/vosa/index.php}} \citep[\texttt{VOSA};][]{Bayo08}. This is a web-based VO-compliant tool developed by the Spanish Virtual Observatory designed to query accessible photometric catalogues and to obtain stellar physical parameters (e.g. temperature, bolometric flux, and mass-loss rate) of the studied sources by fitting theoretical models to the observational SEDs.

\subsection{Photometry compilation and SED construction}
\label{sec:phot}

First of all, we compiled available photometric data from the ultraviolet to the far-infrared in order to obtain a well-covered SED.  To do this, we uploaded the AllWISE, 2MASS, and \G\ photometry collected in the previous step, along with the distance estimations, if available, to \texttt{VOSA}, and queried the VO for additional photometric information from catalogues available through the application. 

We retrieved infrared photometry from the following catalogues: the IRAS catalogue of Point Sources (version 2.0) and the IRAS Faint Source catalogue \citep{Beichman88}; the AKARI/FIS All-Sky Point Source Catalogue \citep{Yamamura10} and the AKARI/IRC Mid-Infrared All-Sky Survey \citep{Ishihara10}; and the MSX6C Infrared Point Source catalogue \citep[version 2.3;][]{Egan03a}. In the latter case, we discarded bands B1 and B2 because the quality of the photometry is low.

We also queried the Deep Near-Infrared Survey \citep[DENIS;][]{Epchtein97}, as well as the VISTA Variables in the Via Lactea Survey \citep[VVV;][]{Saito12} and the VISTA Hemisphere Survey  \citep[VHS;][]{McMahon13}. To avoid fitting issues because our sources are variable in the near-infrared, we only used the VISTA $Z$ and $Y$ bands and the DENIS I band, and we omited the $J$, $H$, and $K_s$ bands in all these cases because they overlap with the 2MASS photometry.

In addition to the optical bands from DENIS and the VISTA surveys, we also retrieved photometry from the PanSTARRS~1 Survey \citep{Chambers17}, which provides photometry in the $g$, $r$, $i$, $z$, and $y$ bands. For a limited number of sources, ultraviolet photometry from the Galaxy Evolution Explorer (GALEX) was also available \citep{Bianchi00}. 

Bad data and upper limits are already flagged by \texttt{VOSA} based on the information provided by the catalogues. As a general rule, we accepted this information as it was provided by the tool. However, in the few cases in which the visual inspection of the SEDs suggested that a data point that was not considered bad was problematic, we manually flagged it as such. 

\subsection{SED shapes} 
\label{sec:shapes}

\begin{figure*}
\centering
        \includegraphics[width=0.85\textwidth] {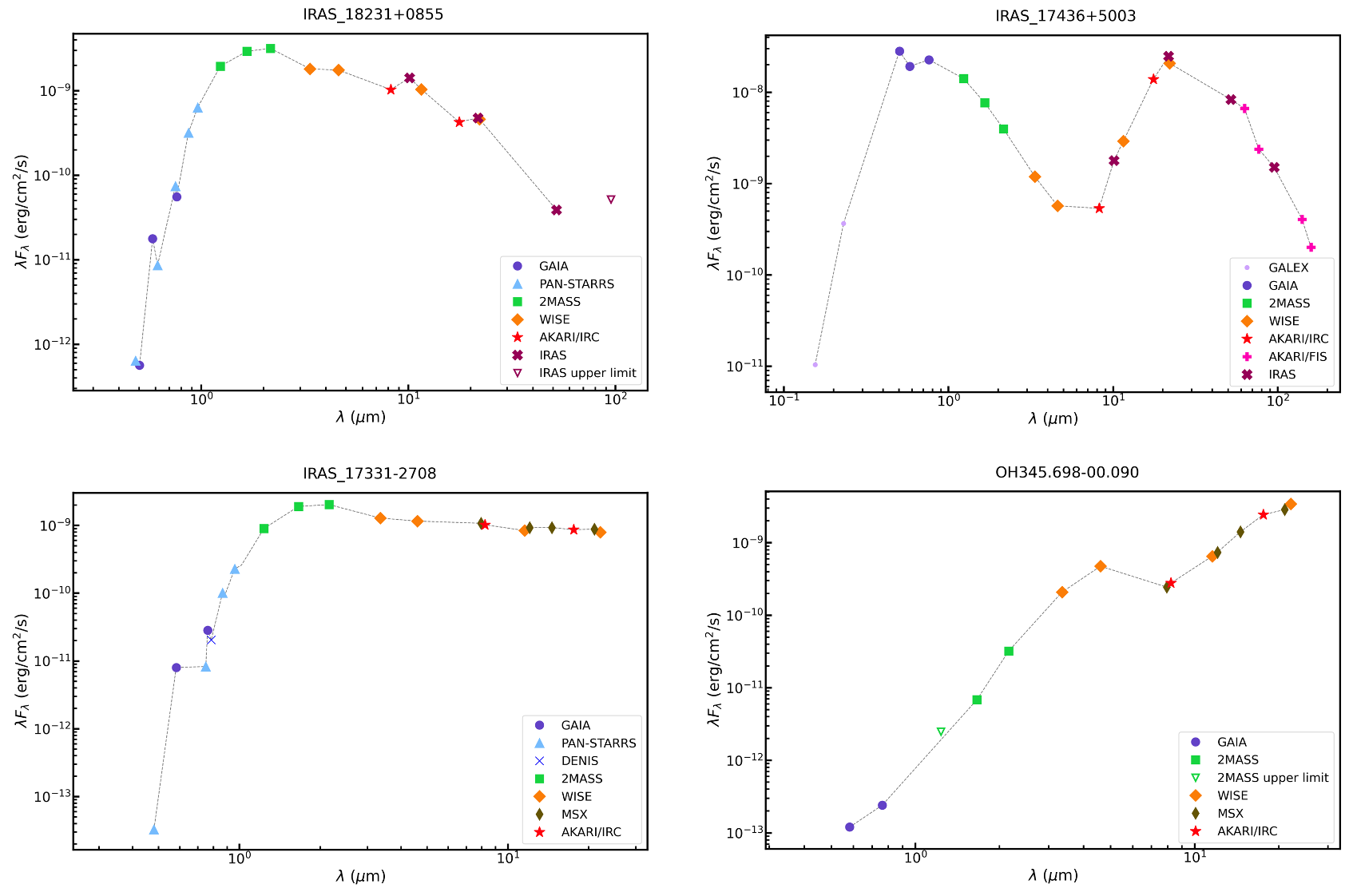}
    \caption{Examples of the different types of SEDs found in the \G\  sample. The photometry we used is listed in the insets of the panels, and the solid line connects the data points to guide the eye. From left to right and from top to bottom: Single-peaked SED, double-peaked SED, flat SED, and peculiar SED. According to the literature, the plotted sources are an OH/IR star, a post-AGB star, another OH/IR star, and a YSO, respectively. An SED graph for any source in the final catalogue can be retrieved through the online service (see Appendix~\ref{sec:Append}).
    }
    \label{fig:shapes}
\end{figure*}

The SEDs constructed with the help of \texttt{VOSA} were visually inspected and classified into four main types based on their appearance in a plot of $\lambda F_\lambda$ versus wavelength (see Fig.~\ref{fig:shapes} for some examples). We describe them below.

\begin{itemize}
     
    \item 
`Single-peaked SEDs' display a clear maximum in the infrared domain. These SEDs correspond to the shape expected for typical AGB stars. The observed distribution is a superposition of emission from the stellar photosphere and the dust in the CSE.
  
    \item 
`Double-peaked SEDs' have two local maxima, one maximum in the mid-infrared, and the other in the optical to near-infrared domain. These two components are identified with the circumstellar shell and the central object, respectively. This type of SED is characteristic of the post-AGB phase, when the shell is being cleared and the central object is already emerging.

    \item 
`Flat SEDs' present a rather flat distribution in the mid- to far-infrared domain, typically with a steep rise at shorter wavelengths (near-infrared and optical), and no flux drop at the longest wavelengths. They might be a borderline case between the two types described above, or the flat appearance might just be caused by a lack of coverage at the longest wavelengths, where the drop is expected to occur.

    \item 
`Peculiar SEDs' have shapes that do not match any of the previous descriptions. Some of these sources seem to be interlopers (sources with OH masers that are not evolved stars, mainly YSOs), while in other cases, the peculiar shape may be the result of associating one or more incorrect counterpart in the photometric catalogues. It may also be caused by poor-quality data. 

\end{itemize}

The SEDs of about 88\% of our sources are classified as single-peaked or flat, and the objects were thus considered AGB star candidates. The next step was to compare them with the predictions from theoretical models.

\subsection{Model fitting}
\label{sec:fits}

The goals of this step were to confirm the AGB nature of the sources and to estimate their bolometric fluxes, and when a distance value was available, also their absolute luminosities. 

For each parameter combination in the chosen model grid, \texttt{VOSA} computes the synthetic photometry for the different filters used to construct the SED, building a synthetic SED that is compared with the observed SED. The best-fitting model is determined by the smallest reduced ${\chi}^2$, although users can choose a different model if they consider it more appropriate. In the calculation of the reduced ${\chi}^2$ for each model of the grid, \texttt{VOSA} weights the photometric data with the inverse of the square of the photometric error.

In our case, the theoretical models used for the fits were the Grid of Red supergiant and Asymptotic giant branch star ModelS (GRAMS) for oxygen-rich stars \citep{Sargent11}. This model grid explores the following parameters: stellar effective temperature from 2100 K to 4700~K; dust shell inner radii of 3, 7, 11, and 15 R$_*$; and 10.0~$\mu$m optical depth from $10^{-4}$ to 26. We set no constraints in this parameter space because were only interested in deriving bolometric fluxes and luminosities, and we therefore set no constraints in this parameter space. 

 Interstellar extinction corrections were applied to sources with a good parallax determination. The value of the interstellar extinction was obtained from the 3D map of Galactic interstellar dust from \cite{Lallement22}. These authors used Gaia EDR3 photometric data combined with 2MASS measurements to derive the extinction density at $\lambda$\,=\,5500\,\AA\ towards any direction in the sky in a volume around the Sun of 6$\times$6$\times$0.8\,kpc³. Knowing the distances to the sources, it is thus possible to use these data to estimate the individual integrated extinction in the line of sight for each of them. 

About 30\% of the sources with a good parallax fall outside the volume that is covered by the extinction map. All these sources are far from the Galactic plane, with Galactic heights larger than $\sim$350\,pc. In these cases, we considered the maximum line-of-sight integrated extinction provided by the map as the interstellar extinction value for the source. Since most of the interstellar dust is located close to the Galactic plane, the cumulative extinction above or below 350-400\,pc from the Galactic plane is probably low, and the effect on the overall SED is negligible.  

The obtained extinction values were then used in \texttt{VOSA} to deredden the observed photometry before model fitting and bolometric flux estimation. \texttt{VOSA} uses the extinction law by \cite{Fitzpatrick99} improved by \cite{Indebetouw05} in the infrared\footnote{\footnotesize \url{http://svo2.cab.inta-csic.es/theory/vosa/helpw4.php?\\otype=star\&action=help\&what=extinctions}}.

The model fitting of an observational SED built from a variety of single-epoch catalogues was a challenging task. First, the agreement between the model and the observations was hampered by the high variability of the AGB star candidates, especially in the optical domain, where the variability amplitude reaches several magnitudes \citep[see e.g. Fig. 9 of][]{Jimenez-Esteban21}. In addition, the longest-wavelength bands (beyond 60~$\mu$m) were in many cases affected by contamination. To avoid that the the fitting results were dominated too much by these effects, we decided to run the fits only in the range between 1 and 60~$\mu$m, so that the data points outside this range were not included in any calculations. 

On the other hand, the photometry data provided by the different catalogues also have very different absolute errors. The smallest errors usually corresponded to the shortest wavelengths, and therefore, these photometric points had the heaviest weights in the fitting process. The consequence was that in many cases, the NIR data are very well fitted, while the mid- and far-infrared domains are completely discrepant from the best-fitting model. Thus, after several tests, we decided to upload the photometry onto \texttt{VOSA} with zero errors for all bands. When this is done, \texttt{VOSA} considers an error of 10\% of the corresponding flux in all cases. This is still lower than the typical variability amplitude of our sources, but better than the usually much smaller absolute photometric errors provided by the catalogues, especially in the near-infrared range.

The fits run with these conditions (zero input errors, and only in the wavelength range between 1 and 60~$\mu$m) agreed better overall in the central part of the SED (which contains the peak of the distribution in the case of a single-peaked SED) than the fits performed over the entire range and/or considering absolute photometric errors. Because \texttt{VOSA} computes the bolometric flux using the observed photometry whenever possible and it only interpolates with the model when no observations are available, this ensured that the errors introduced in this computation through variability or contamination were minimised.

All fits were visually inspected to ensure their reliability. When the SED clearly deviated from the model, we considered it a poor fit. Based on this inspection, the majority of SEDs classified as single-peaked (92\%) turned out to have relatively good fitting results. A fraction of the flat SEDs also displayed good fits; for some of these sources, the models suggested that a flux drop indeed occurred beyond the range covered by the observations. As expected, all double-peaked and peculiar SEDs had poor fits.

For nine sources in the sample, the use of the restricted wavelength range had the consequence that the number of available photometric points (excluding upper limits) was too low to perform the fit. All of them displayed single-peaked SEDs, except for one, which had a peculiar SED.

%
\begin{table*}
\caption{ \G\ sample of OH/IR stars. 
}
\label{tab:OHIR}  
\footnotesize
\begin{center}
\begin{tabular}{l l l}     
\hline\hline       
Column & Units & Description\\ 
\hline  
  RA\_(ICRS) & deg &  \G-DR3 right ascension (Epoch J2016) \\
  DEC\_(ICRS) & deg & \G-DR3 declination (Epoch J2016) \\ 
  \G\_DR3\_ID  & & \G-DR3 unique source identifier \\ 
  SIMBAD\_ID && SIMBAD source identifier \\
  WISE\_ID && WISE source identifier \\
  2MASS\_ID && 2MASS source identifier \\ 
  G & mag & \G-DR3 $G$ band mean magnitude \\    
  BP & mag & \G-DR3 integrated $G_{BP}$ band mean magnitude \\  RP & mag & \G-DR3 integrated $G_{RP}$ band mean magnitude \\  parallax & mas & \G-DR3 parallax \\    
  parallax\_error & mas & \G-DR3 standard error of parallax \\ 
  distance$^{*}$ & pc & Geometric distance estimated by \cite{Bailer-Jones21} (see text) \\
  A$_v$$^{*}$ & mag & Interstellar extinction estimated by \cite{Lallement22} (see text) \\
  Fbol & erg/cm$^2$/s & Bolometric flux obtained from the SED fitting (see text) \\
  Luminosity$^{*}$ & \Lsun & Stellar luminosity obtained from the bolometric flux and the distance\\
\hline
\end{tabular}
\end{center}
Note: The full table is available electronically at the CDS. $^{*}$For objects with a good parallax (see text).\\
\end{table*}
%

%
\begin{table*}
\caption{Sources with \G-DR3 counterparts that were excluded from the \G\ sample of OH/IR stars. }\label{tab:rejected}  
\footnotesize
\centering          
\begin{tabular}{l l l}     
\hline\hline       
Column & Units & Description\\ 
\hline  
  RA\_(ICRS) & deg &  \G-DR3 right ascension (Epoch J2016) \\
  DEC\_(ICRS) & deg & \G-DR3 declination (Epoch J2016) \\ 
  \G\_DR3\_ID  & & \G-DR3 unique source identifier \\ 
  SIMBAD\_ID && SIMBAD source identifier \\
  WISE\_ID && WISE source identifier \\
  2MASS\_ID && 2MASS source identifier \\ 
  SED\_type & & SED shape type: single-peaked (SP), double-peaked (DP), flat (F) or peculiar (P)\\
  Fit & & Flag indicating if the fit was good (G) or bad (B) \\
  SIMBAD\_class & & Main SIMBAD classification\\
  Rejection & & Reason for rejection \\
\hline                  
\end{tabular}
\tablefoot{The full table is available electronically at the CDS.}
\end{table*}
%

\section{The \G\ OH/IR star sample}
\label{sec:cat}

After the SED inspection, the fits for 1190 sources were considered good. This is nearly 80\% of the \G\ sample obtained by the cross-match process described in Sect.~\ref{sec:sample}. Although these sources are likely OH/IR stars, a good fit does not guarantee that other types of sources are completely removed from the sample. (To reiterate: We use the designation `OH/IR star' as a general term to refer to AGB stars with OH maser emission). Thus, in order to check the reliability of the identification of OH/IR stars and to remove interlopers from the sample, we cross-matched the whole \G\ sample with the SIMBAD database.

SIMBAD had counterparts for 99.5\% of the \G\ sample (1481 sources), the classification of most of which was compatible with an OH/IR star. The largest number of interlopers were post-AGB stars (49 sources), followed by 19 red supergiant stars (RSGs), 13 planetary nebulae, 13 YSOs, two interstellar medium sources, and three multiple stellar systems. In total, the classification of 99 sources (6.7\% of the sources with SIMBAD counterparts) was not compatible with an OH/IR star. Of these, only 15 had a good model fit, namely all the RSGs and one post-AGB star. This is $\sim$1.3\% of the sources in the \G\ sample with a good fit. Thus, the expected contamination of the sources with good model fits is very low ($\sim$1-2\%).

Consequently, we defined the `\G\ OH/IR star sample' as sources with a good model fit and with a SIMBAD classification compatible with an AGB star.
In total, 1172 sources (79\%) out of the 1487 members of the \G\ sample were assigned to this sample. Of these, 222 had Bailer-Jones geometric distances and thus luminosity estimates.

We present these 1172 sources in an online table (Table\,\ref{tab:OHIR}), along with their \G-DR3, SIMBAD, AllWISE, and 2MASS identifications, their \G\ coordinates, brightnesses, and parallaxes, their geometric distances from \citet{Bailer-Jones21}, their derived bolometric fluxes and, when distances are available, their luminosities from the SED fitting. In the remainder of the paper, we focus on this sample of stars. This table, along with the observational photometry data we used to build and analyse the SEDs, is also electronically available in an online archive (see Appendix \ref{sec:Append}) and at the CDS VizieR catalogue service.\footnote{\footnotesize \url{https://vizier.cds.unistra.fr/viz-bin/VizieR}} Additionally, the sources that were excluded from the \G\ OH/IR star sample are presented in a second online table (Table\,\ref{tab:rejected}), along with their \G-DR3, SIMBAD, WISE, and 2MASS identifications, \G\ coordinates, their SED type, the reliability of the model fit (good or poor) as assessed by eye, the SIMBAD classification when available, and the reason for rejection. 

\begin{figure}
\centering
        \includegraphics[width=0.97\columnwidth]{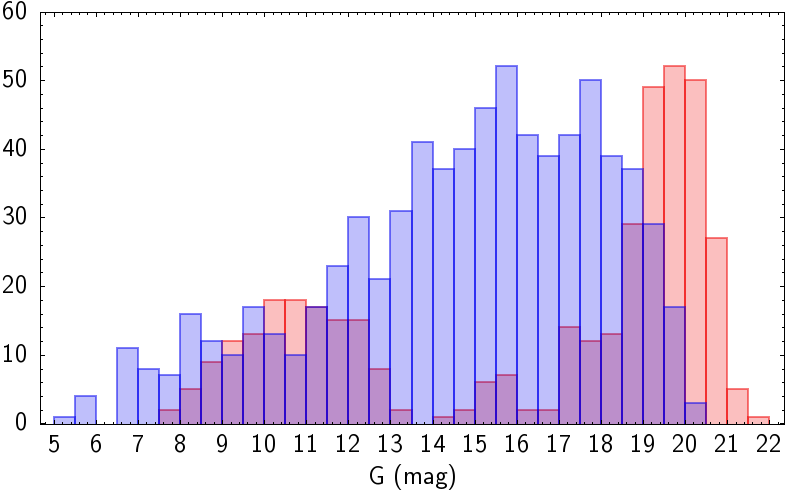}
    \caption{Objects of the \G\ OH/IR star sample that are included and missed (blue and red histogram, respectively) in the second \G\ catalogue of LPV candidates \citep[]{Lebzelter23}.}
    \label{fig:Lebzelter23-cat}
\end{figure}

The \G\ OH/IR star sample should consist mostly of long-period variable stars and consequently is expected to be contained in the second \G\ catalogue of LPV candidates \citep{Lebzelter23}. However, only 475 (40\%) OH/IR stars could be recovered in the LPV catalogue. The recovery rate is strongly dependent on the optical brightness. Figure \ref{fig:Lebzelter23-cat} shows the number of OH/IR stars that we recovered as function of G-band brightness. The number of OH/IR stars included in the LPV catalogue decreases strongly for objects with G$\ge17$ mag. Optically bright OH/IR stars G$\le13$ mag are also frequently missed. \cite{Lebzelter23} obtained recovery
rates $\ge80$\% relative to the LPV published by ASAS-SN and OGLE-III independent of \G\ brightnesses. We found high recovery rates like this relative to the \G\ OH/IR star sample only in the brightness range $13 \le$ G $\le 17$. 

We also cross-matched our sample with the sample of nearly 66,000 Mira variables included in the OGLE Collection of Variable Stars \citep{iwanek22}. We found 345 sources in common with this catalogue, which is nearly 30\% of the Gaia OH/IR sample. The majority of the remaining sources in our sample are located outside the overlapping area between the two catalogues. Most Gaia OH/IR sources within the overlap region that lack a match in the OGLE catalogue are brighter than G$\sim$12 mag, and they lie outside the brightness range of the OGLE survey.

\section{Properties of the \G\ sample of OH/IR stars}
\label{sec:properties}

In this section, we discuss the main properties of the \G\ OH/IR star sample. A more detailed scientific analysis is deferred until the remaining samples (carbon-rich AGB stars and oxygen-rich AGB stars without OH masers) are presented in forthcoming papers.

\subsection{Spatial distribution}
\label{sec:sky}

\begin{figure*}
\centering
        \includegraphics[width=0.8\textwidth]{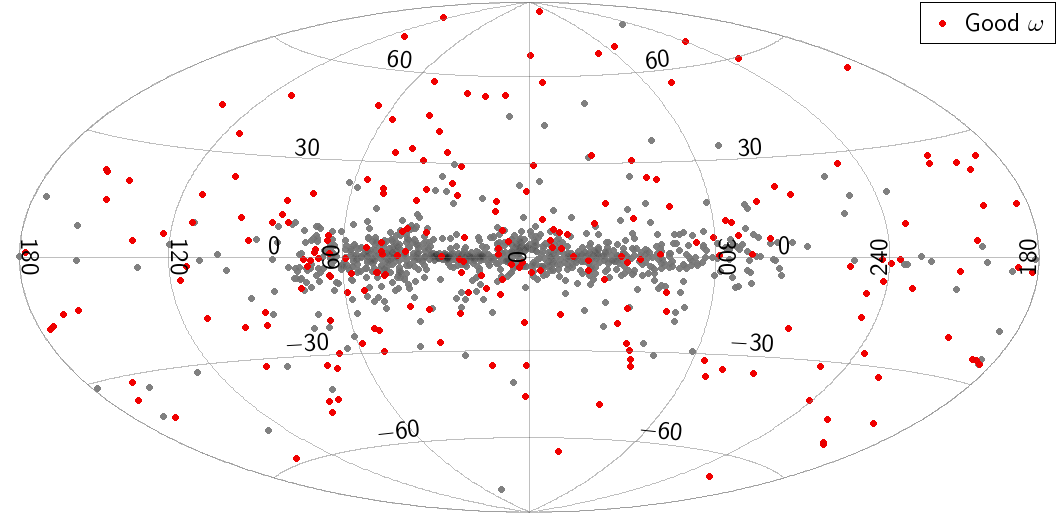}
    \caption{Sky distribution of the \G\ sample of OH/IR stars. The sources with good parallaxes (relative parallax error $<20$\% and astrometric excess noise $< 1.5$~mas) are shown in red, and the rest are plotted in grey.} 
    \label{fig:sky}
\end{figure*}

Figure~\ref{fig:sky} shows the distribution in the sky of the \G\ OH/IR star sample. As explained in Sect.~\ref{sec:Gaia-xm}, an area of two degrees around the Galactic centre was removed from the sample due to crowding issues.

The sources are spread throughout the sky, although there is a concentration toward the inner Galactic plane. This was previously reported for known Galactic O-rich AGB stars \citep[and references therein]{Ishihara11}. However, it is important to recall that this sky distribution may be biased because it originates from a compilation of OH maser sources from the literature, which is dominated by some surveys that were made along the Galactic plane \citep{Engels15b}.

\begin{figure}
\centering
        \includegraphics[width=0.97\columnwidth]{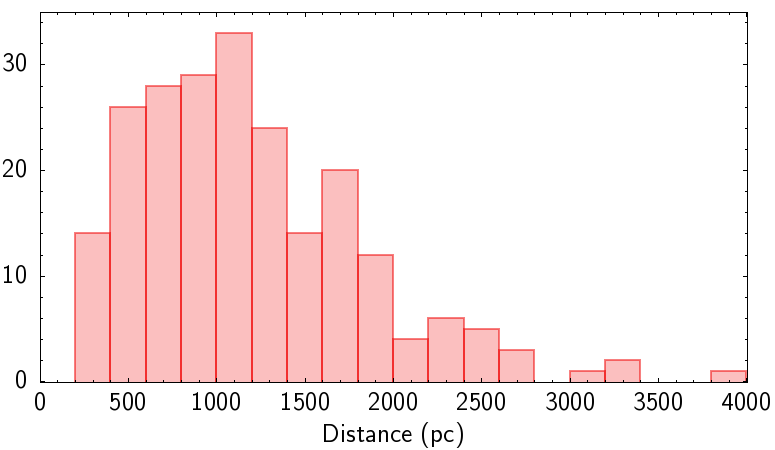}
     \caption{Histogram showing the geometric distances from \citet{Bailer-Jones21} for the OH/IR stars with good parallax measurements. }
    \label{fig:dist}
\end{figure}

Figure~\ref{fig:dist} shows the distribution of \citet{Bailer-Jones21} geometric distances for the 222 OH/IR stars with good parallaxes. These stars are located within 4.0 kpc, and only $\sim$10\% of them are located beyond 2\,kpc. The mean and median distance of this sample are 1.29 and 1.17~kpc, respectively. 

These sources are also shown in Fig.\,\ref{fig:sky} in red. They are less concentrated towards the inner Galactic plane than the whole OH/IR sample. This suggests that this is a local sample of stars.

\begin{figure}
\centering
        \includegraphics[width=0.8\columnwidth]{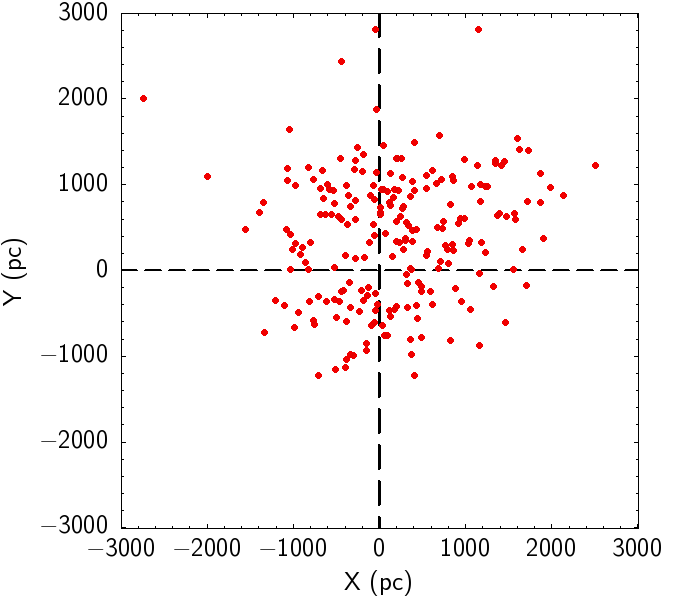}
    \caption{Spatial distribution of the OH/IR stars with good parallax projected onto the Galactic plane. The Sun is located at the centre (0,0), and the direction towards the Galactic centre is upwards along the vertical dashed line.}
    \label{fig:GalPlane}
\end{figure}

Figure\,\ref{fig:GalPlane} shows the relative position with respect to the Sun of the OH/IR stars with good parallax and \citet{Bailer-Jones21} distances. The sample is more numerous toward the inner Galaxy, and it is confined to distances $\le1.1$ kpc in the outer Galaxy. There also seems to be some east-west asymmetry. This projected spatial distribution is very similar to the distribution found by \cite{Ishihara11} for O-rich AGB stars in the Galactic plane for the distance range covered by our sample. Thus, although we cannot completely rule out with the current data that these features are caused by selection effects, they seem to be real.

\begin{figure}
\centering
        \includegraphics[width=0.97\columnwidth]{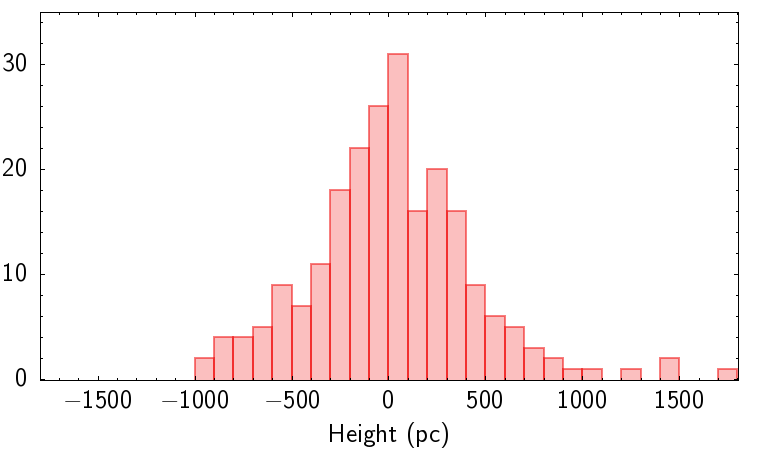}
     \caption{Galactic height distribution of OH/IR sources with geometric distances from \citet{Bailer-Jones21}.} 
    \label{fig:GalHeight}
\end{figure}

Finally, we show in Fig.\,\ref{fig:GalHeight} the distribution of height above the Galactic plane of the OH/IR sources with \citet{Bailer-Jones21} distances. In this case, the distribution is very symmetric with respect to the Galactic plane, and the number of sources decreases with height. 

To summarise, the group of OH/IR stars with available distances trace a more or less local sample of OH/IR stars that is somewhat more numerous in the direction of the inner part of the Galaxy and is distributed roughly symmetrically with respect to the Galactic plane. It is expected that the OH/IR stars with a poor parallax determination are mainly located in the direction of the inner galaxy and at farther distances.

\subsection{Magnitudes and colours}
\label{sec:mag_col}

We now discuss the magnitude and colour distributions of the OH/IR stars with \G\ counterparts. Only sources with good photometry are considered, that is, a photometry quality flag A, B, or C for WISE photometry; a photometry quality flag A, B, C, or D for 2MASS photometry; and a relative error lower than 10\% for the \G\ fluxes. Furthermore, all sources with BP$>$20.5 mag were removed from the magnitude and colour distributions using this filter because of a known bias that produces systematically unreal blue colours for faint red sources \citep{Fabricius21}. 

\begin{figure*}
\centering
        \includegraphics[width=0.97\columnwidth]{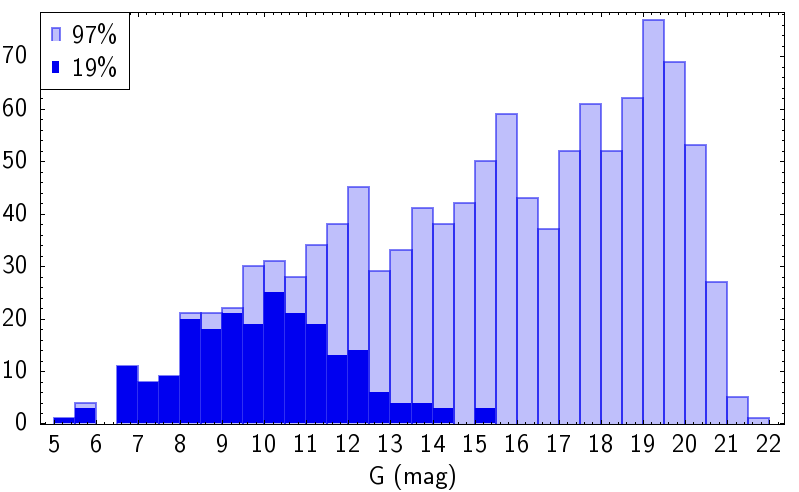}
        \includegraphics[width=0.97\columnwidth]{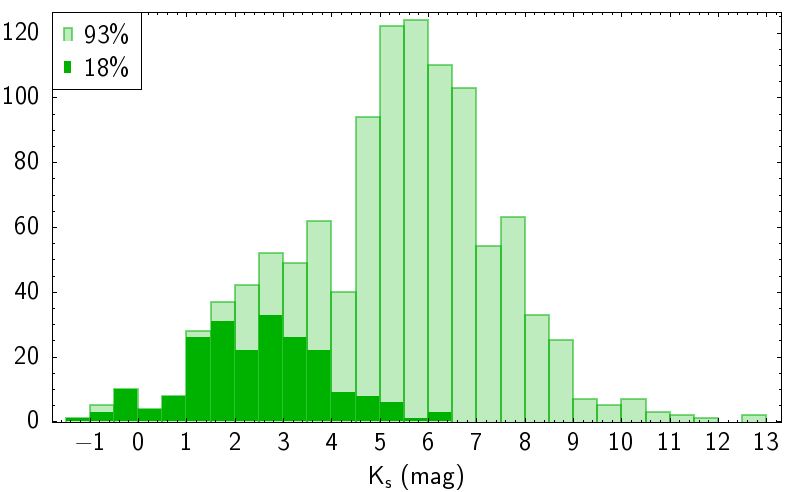}
        \includegraphics[width=0.98\columnwidth]{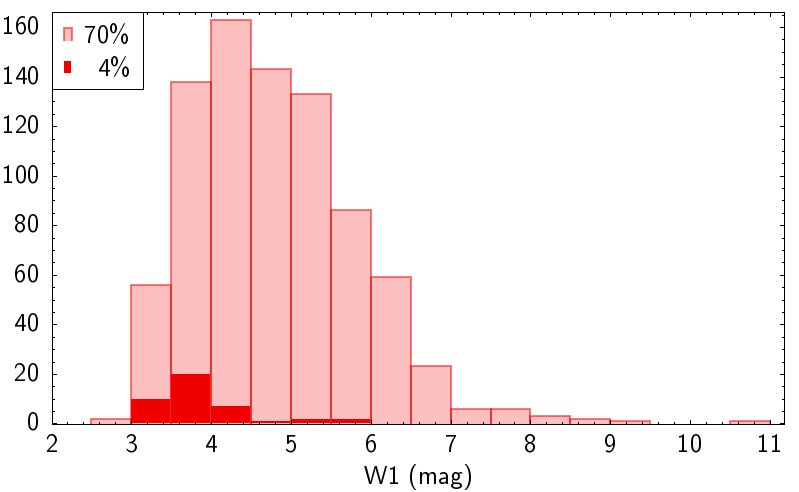}
        \includegraphics[width=0.97\columnwidth]{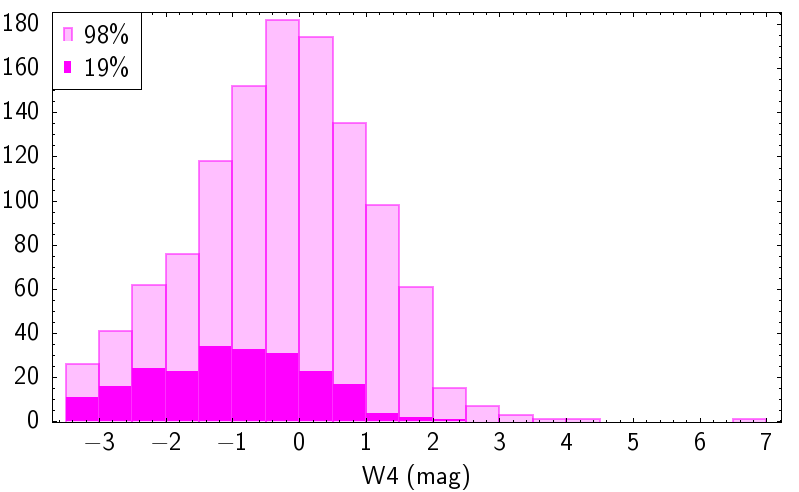}
    \caption{Magnitude distributions of the \G\ OH/IR sample (semi-filled histograms) and of the subsample of sources with good astrometry (filled histograms). From left to right and from top to bottom: \G\ G band, 2MASS $K_s$ band, and the WISE W1 and W4 bands. The percentages refer to the fraction of sources with respect to the total \G\ OH/IR sample of 1172 sources.}
    \label{fig:mags}
\end{figure*}

Figure\,\ref{fig:mags} shows the magnitude distribution of the whole OH/IR sample (semi-filled histograms) and of the sources with good parallaxes (filled histograms). Four bands are displayed, namely the \G\ G-band, 2MASS $K_s$, and the WISE W1 and W4 bands. In the G band, the number of detected sources increases towards higher magnitudes. The drop at the highest observed magnitudes is most likely due to incompleteness. Most sources with G$<10$~mag also have good parallaxes. This is expected because brighter sources should have better quality detections and better \G\ astrometric parameters.
On the other hand, the W4 magnitude distribution of the whole sample and of the good-parallax sample are similar. This suggests that there is no intrinsic difference between the sources with a good parallax and the rest, except that they are probably closer and less affected by interstellar extinction.

\begin{figure*}
\centering
        \includegraphics[width=0.97\columnwidth]{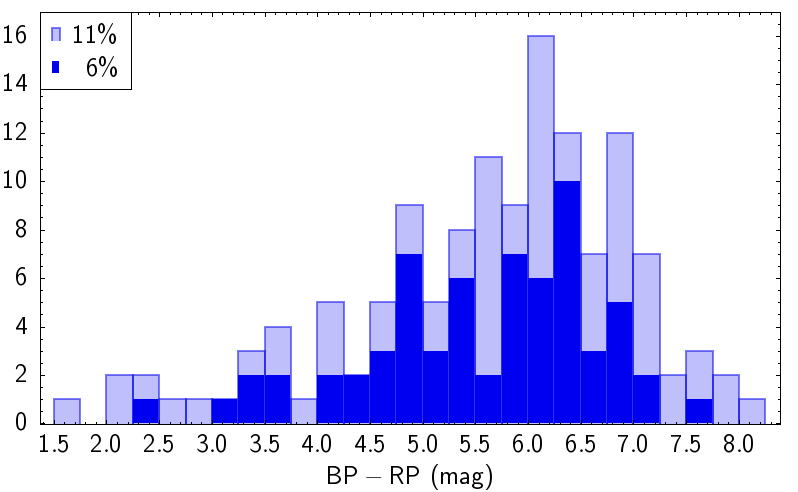}
        \includegraphics[width=0.97\columnwidth]{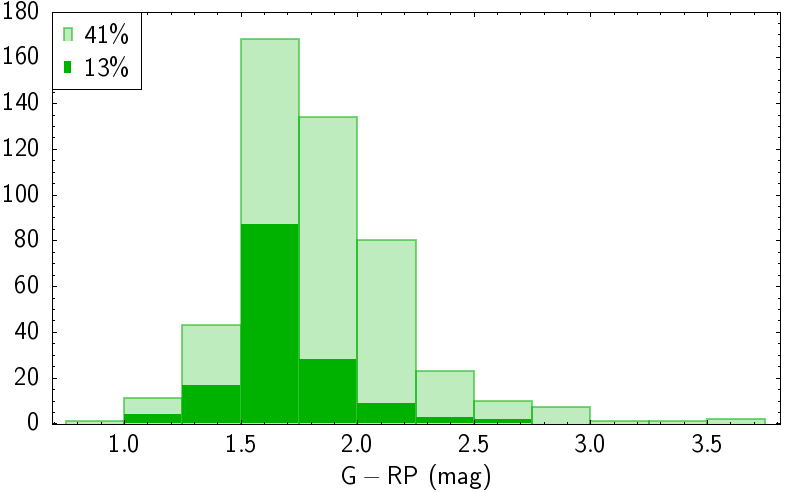}
        \includegraphics[width=0.97\columnwidth]{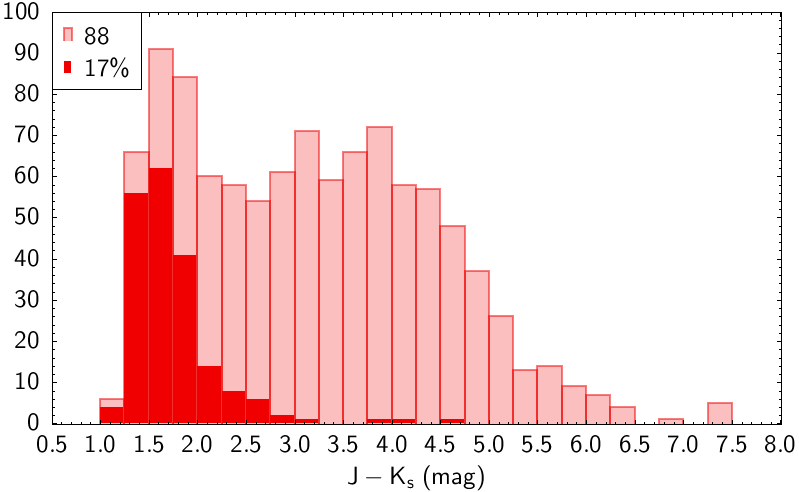}
        \includegraphics[width=0.97\columnwidth]{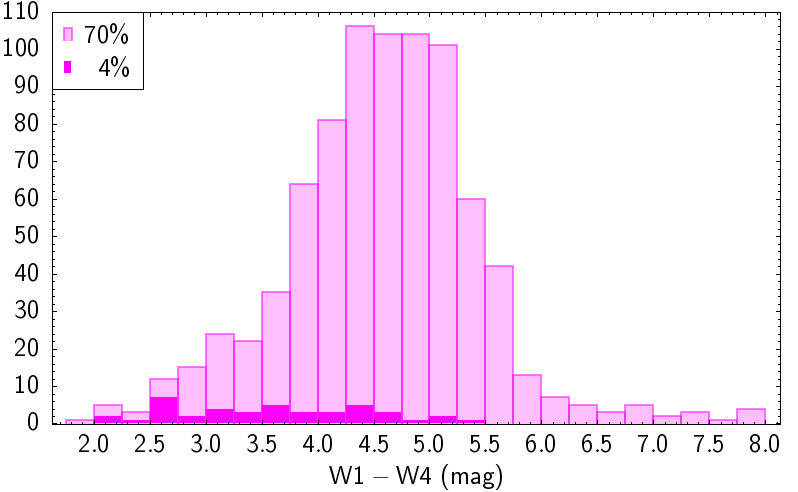}
    \caption{Colour distributions of of the \G\ OH/IR sample (semi-filled histograms) and of the subsample of sources with a good astrometry (filled histograms). From left to right and from top to bottom: \G\ BP--RP and G--RP, 2MASS $J-K_s$, and WISE W1$-$W4 colours. The percentages refer to the fraction of sources with respect to the total \G\ OH/IR sample of 1172 sources. }
    \label{fig:colours}
\end{figure*}

Figure~\ref{fig:colours} shows the distributions of two \G\ colours (BP$-$RP and G$-$RP), as well as two infrared colours ($J-K_s$ and W1$-$W4), for the likely OH/IR stars with \G\ counterparts (semi-filled histograms) and of the subsample of sources with a good astrometry (filled histograms). 

As expected, the sources tend to have quite red optical colours, as is evident from the BP$-$RP colour distribution. This colour is available for only 11\% of the \G\ detected OH/IR stars because many faint sources with red colours were only detected by \G\ in the RP band, and as explained above, sources with BP magnitudes $>$20.5 mag were removed. 
The plot appears to be truncated at BP$-$RP\,$\sim 8.3$ simply because of the \G\ detection capabilities; the OH/IR that are left undetected are likely to extend the colour range to even higher values.

The BP$-$RP distribution of the \G\ OH/IR sample peaks at rather red colours BP$-$RP$\sim6$ mag. This is different from the colour distributions found for \G\ LPV in general. \cite{Mowlavi18} for DR2 and \cite{Lebzelter23} for DR3 considered LPV with G-band amplitudes QR5(G)$>$1 mag predominantly consisting of Mira variables, which make up $10-20$\% of all the LPVs that were identified with QR5(G)$>$0.2 mag. In amplitude-colour diagrams, this group of Mira variable candidates stands out and has colours BP$-$RP$>4$ mag \citep{Mowlavi18}. The fraction of stars with BP$-$RP$>4$ in our sample is $\sim$85\%, indicating that the small amplitude QR5(G)$<$1 mag variables are underrepresented compared to the general LPV population seen by \G.

The good-parallax sources are not at all different in this respect. However, they are among the bluest sources in the near-infrared, as shown in the $J-K_s$ colour distribution. This again suggests a lower extinction than for the whole sample.

\subsection{Bolometric fluxes and luminosities}
\label{sec:lum}

\begin{figure*}
\centering
        \includegraphics[width=0.97\columnwidth]{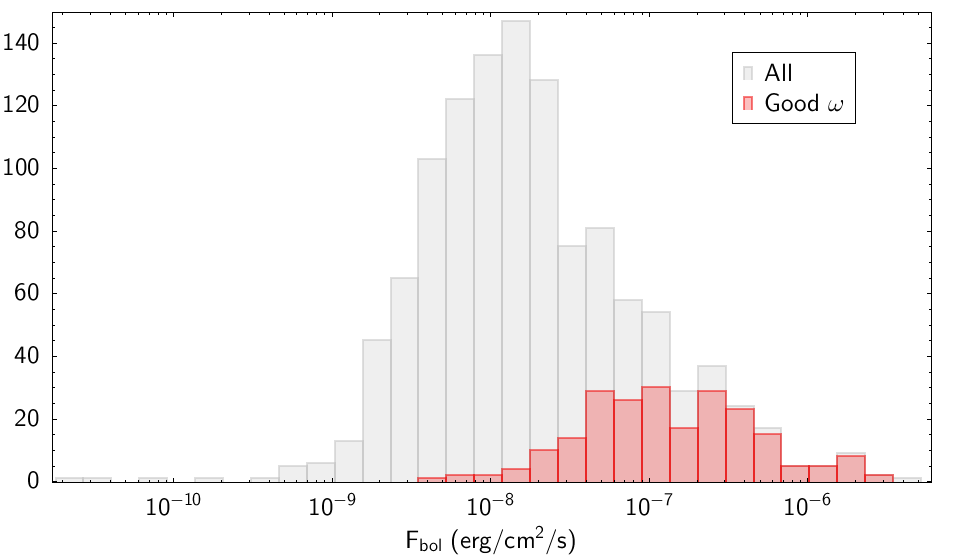}
        \includegraphics[width=0.97\columnwidth]{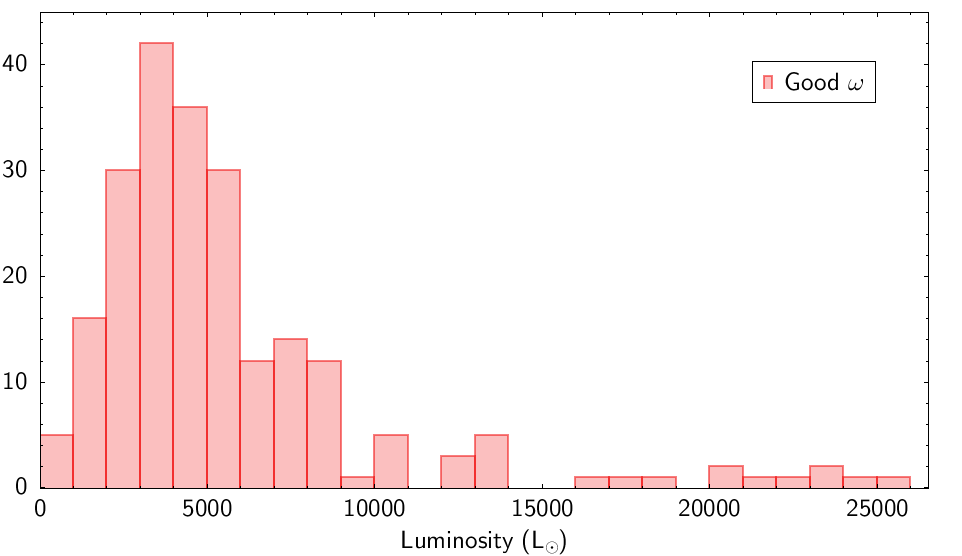}
    \caption{Left panel: Distribution of bolometric fluxes (left) of the \G\ OH/IR sample. The grey and red histograms correspond to the whole sample and to stars with good parallaxes (good $\omega$), respectively. Right panel: Luminosity distribution for the subsample with good parallaxes.}
    \label{fig:flux_lum}
\end{figure*}

The left panel of Fig.~\ref{fig:flux_lum} shows the distribution of bolometric fluxes obtained from the SED fitting for our OH/IR sample. The stars with good parallaxes are among the brightest objects, as expected.

The luminosity of the sources in this latter subsample was computed by VOSA from the bolometric flux and the Bailer-Jones geometric distances. We recall that the bolometric flux was obtained after correcting the observed photometry for interstellar extinction (see Sect.\ref{sec:fits}). This distribution is shown in the right panel of Fig.~\ref{fig:flux_lum}. Most luminosity values are as expected for AGB stars, typically between 1,000 and 10,000\,L$_{\odot}$, with an extended tail up to $\sim$25,000\,L$_{\odot}$. The peak of the distribution is around 4000\,L$_{\odot}$.

The luminosities of five sources are lower than 1000~L$_{\odot}$: \object{IRAS~17528+1144}, \object{IRAS~18042--2517}, \object{IRAS~20234--1357}, \object{SY Sco}, and \object{R Vir}. All but IRAS~20234--1357 are in the \G-DR3 catalogue of LPV stars \citep{Lebzelter23}, and their \G\ light curves show clear Mira-like variability, with periods ranging from $\sim$150 to $\sim$600\,d. The remaining source is included in the ASAS-SN catalogue of variable stars \citep{Jayasinghe18}. It also shows Mira-like variability with a period of $\sim$550\,d. Hence, all five sources therefore appear to be genuine AGB stars. Most likely, their distances are underestimated, and so are their luminosities.

\subsection{\G\ Hertzsprung-Russell diagram}
\label{sec:hrdiag}

\begin{figure*}
\centering
        \includegraphics[width=0.97\columnwidth]{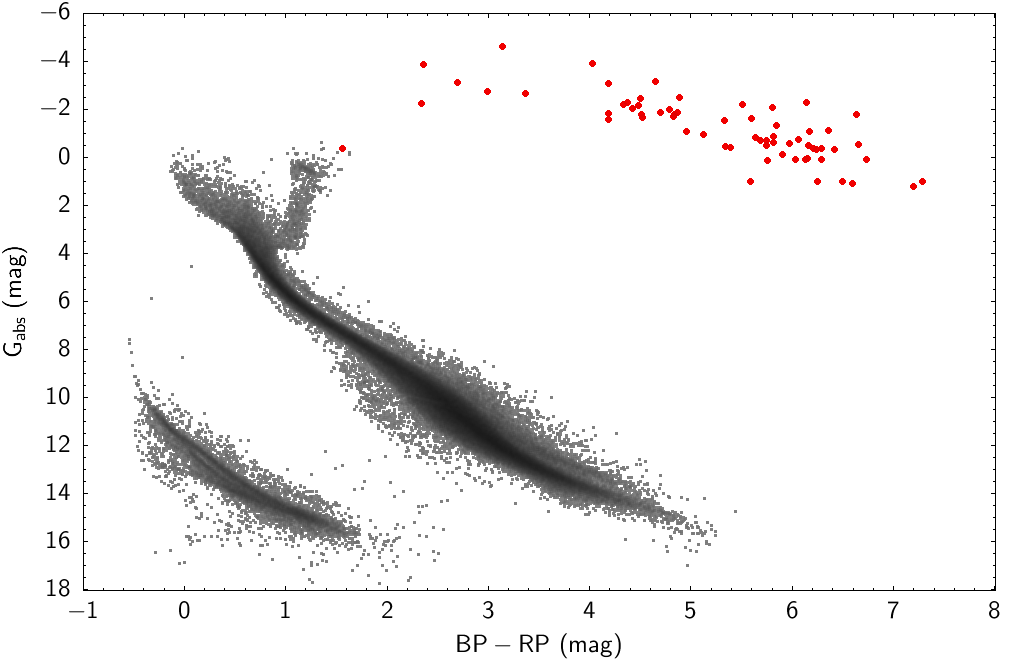}
        \includegraphics[width=0.97\columnwidth]{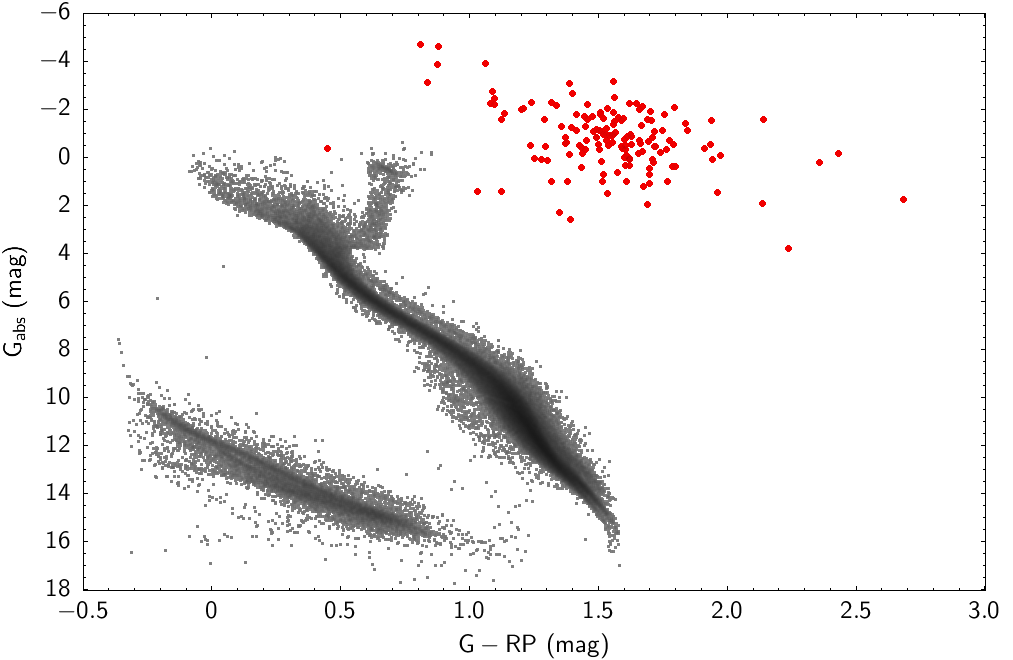}
    \caption{\G\ DR3 HR diagrams using the extinction-corrected BP$-$RP and G$-$RP colours (left and right panel, respectively)
    for the subsample of sources with good parallaxes (red circles), compared to the location of the 100\,pc stellar population (in grey). In each diagram, only sources with good photometry according to the criteria outlined in Sect.~\ref{sec:mag_col} are displayed.
    }
    \label{fig:hrdiag}
\end{figure*}

Using the \citet{Bailer-Jones21} distances and the dereddened \G\ magnitudes, we constructed the Gaia Hertzsprung-Russell (HR) diagram for the OH/IR sample. This diagram displays the G-band absolute magnitude versus one \G\ colour. 

Two versions of this diagram are shown in Fig.~\ref{fig:hrdiag}: In the left panel, we show the commonly used representation with the BP$-$RP colour. Due to the quality criteria for the photometry imposed in Sect.\,\ref{sec:mag_col} and especially because we omitted the \G\ BP photometry for faint red sources, only one-third of the sources with distance estimations can be plotted in this diagram (65 sources). We therefore also show a representation with the G$-$RP colour in the right panel of Fig.~\ref{fig:hrdiag}, where a larger number of sources (N=150) from our OH/IR sample can be displayed. In both diagrams, we also plot a sample of \G\ nearby stars within 100 pc following the same selection criteria as used by \cite{Jimenez-Esteban23}. As the interstellar extinction for these stars is negligible, no dereddening was applied to their \G\ photometry.

In the HR diagram using the BP$-$RP colour, the sources are spread over a wide colour range $2.3 \la ({\rm BP}-{\rm RP}) \la 7.3$ mag. The absolute G-Band magnitude \Gabs\ decreases from \Gabs\ $\sim -4.6$ to  \Gabs\ $\sim +1.2$ with increasing BP$-$RP colour. The locus of the \G\ OH/IR sample in this diagram is similar to the locus of the LPV candidates in the first \G\ catalogue \citep{Mowlavi18}, and this is the expected location for AGB stars \citep{GaiaCollaboration19-Eyer}. 

In \G-selected LPV samples, the bulk of objects has colours $2 \la$\,BP$-$RP\,$\la 4$ \citep{Lebzelter23} and shows semi-regular variability \citep{Gavras23}. The \G\ OH/IR star sample, in contrast, has predominantly redder \G\ colours, which are frequent among large-amplitude periodically pulsating Mira variables \citep{Mowlavi18,Gavras23}. This reflects the larger fraction of OH maser emitters among the AGB stars with higher mass-loss rates and hence redder colours (see the discussion in \cite{Lewis90b} for IRAS colours).

It must be noted that the BP$-$RP colours are averages, and their uncertainty is influenced by the coverage of the light curves by \G\ observations. The colours can vary by 2--3 mag depending on the phase of the pulsation cycle, and together with the brightness variations, they cause loops in the HR diagram \citep{GaiaCollaboration19-Eyer, Mowlavi18}. Because the light curve will be better covered, the location in the diagram for a given star may be different in future GAIA releases, while the area covered by AGB stars in the HR diagram is not expected to change globally considered.

\begin{figure}
\centering
        \includegraphics[width=0.95\columnwidth]{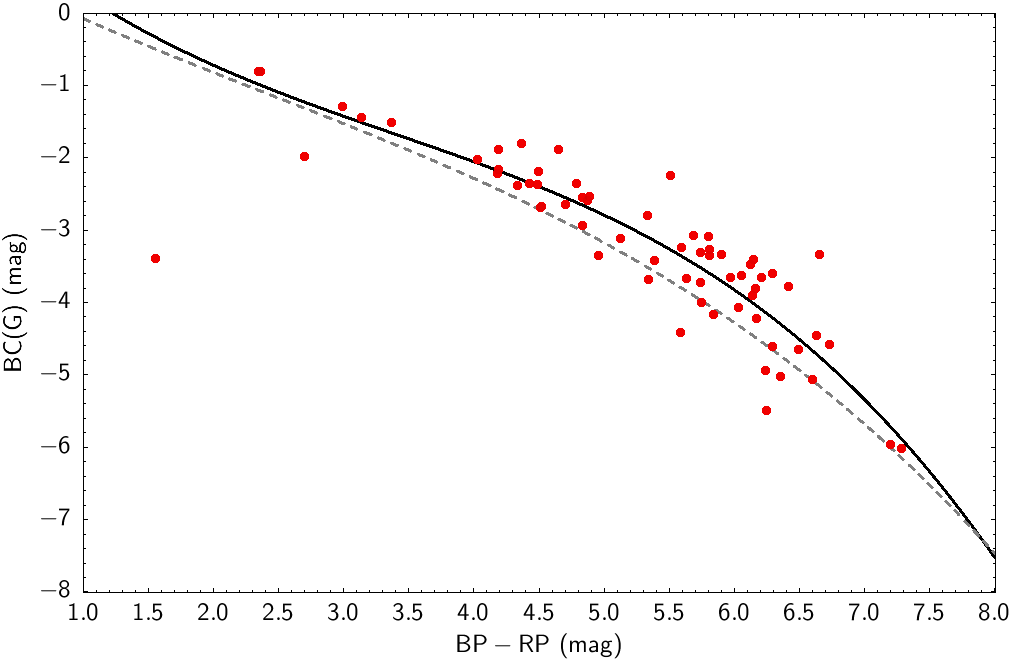}
    \caption{Bolometric corrections $BC(G) = M_{bol} - G_{{abs}}$ for the subsample of sources from the \G\ OH/IR stars with good parallaxes. The outlier in the diagram is IRAS\,18319-1001, as discussed in the text. The solid curve shows the fit to the data as given in Sect. \ref{sec:hrdiag}, and the broken curve is the bolometric correction as a function of BP$-$RP colour proposed by \cite{Lebzelter19} without interstellar extinction.
    }
    \label{fig:bol_corr}
\end{figure}

The increasing intrinsic faintness in the G band with BP$-$RP colour can be attributed to the increasing circumstellar extinction. Quite significant bolometric corrections are needed to infer the stellar luminosities from the G-band absolute magnitudes. This is shown in Fig. \ref{fig:bol_corr}, where the bolometric corrections $BC(G) = M_{bol} - G_{{abs}}$ for the subsample of sources with good parallaxes are given. The bolometric magnitudes $M_{bol}$ were obtained from the luminosities determined in Sect. \ref{sec:lum}.
We fitted the data with a third-degree polynomial, giving
\begin{eqnarray}
BC(G) &=& 1.662 - 1.701\, ({\rm BP} - {\rm RP}) \\\nonumber
    & & + 0.317\, ({\rm BP} - {\rm RP})^2 - 0.031\, ({\rm BP} - {\rm RP} )^3.
\end{eqnarray}
The fit is shown as the solid curve in Fig. \ref{fig:bol_corr}. This fit was previously introduced by \cite{Lebzelter19} using the original \G\ photometry. Their results are shown in the figure as a broken curve. The bolometric correction curve provides a means to obtain stellar luminosities using (dereddened) \G\ photometry alone.
From the scatter around the curve, we estimate an uncertainty of $\Delta M_{bol} = 0.5 (1\sigma)$ mag in the colour range $4 \le {\rm BP} - {\rm RP} \le 7$, which is reasonably well populated by objects.

A few objects are detached from the main locus of our sample in the \G\ HR diagram. The most obvious case is \object{IRAS~18319-1001}, which is placed close to the tip of the red giant branch in the HR diagram using the BP$-$RP colour (G$_{abs}$\,$\approx$\,--0.4\,mag, BP$-$RP\,$\approx$\,1.6\,mag, and G$-$RP\,$\approx$\,0.5\,mag). In this region, type II Cepheid variable stars are also expected \citep{GaiaCollaboration19-Eyer,Rimoldini19}. However, IRAS~18319-1001 is the counterpart of the OH maser \object{G22.136-0.389} that was detected by the recent THOR OH maser survey \citep{Beuther19}, which displays a convincing double-peaked profile typical of an OH/IR star. Inspection of the SED shows that the infrared part is very well fitted with a GRAMS model, in accordance with its classification as an OH/IR star. However, the optical data are brighter than the model, and the brightness toward shorter wavelengths decreases not as strongly. Although the separation between the 2MASS counterpart and the selected \G\ counterpart is about 1\arcsec, the association is almost certainly erroneous, as judged from a detailed comparison of the counterpart positions in Pan-STARRS and 2MASS images. The \G\ optical counterpart is a field star in the line of sight of the apparently optically obscured OH/IR star, which dominates the IR emission. Thus, the peculiar position of IRAS~18319-1001 in Fig. \ref{fig:hrdiag} (and Fig. \ref{fig:bol_corr}) is quite likely due to a misidentification. This case shows that although great care was taken to verify the correct association of the \G\ counterparts, a few cases of contamination with field stars may remain.

In the HR diagram that is based on the G--RP colour (Fig.~\ref{fig:hrdiag}, right panel), the brightness of a group of four sources with  G$_{abs} \ge 1.0$ and G$-$RP\,$<1.5$ appears to be too low for their G$-$RP colour. These are \object{IRAS~11525-5057}, \object{IRAS~17020-5254} (both with G$_{abs}$\,$\sim$\,1.4\,mag and G$-$RP\,$\sim$\,1.1\,mag), and \object{IRC +30292}, \object{IRAS~14591-4438}
(both with G$_{abs}$\,$\sim$\,2.4\,mag and G$-$RP\,$\sim$\,1.3\,mag). \G-DR3 provides epoch photometry for these four sources. All show Mira-like variability in the \G\ G-band light curves. Their periods range from $\sim$\,490 to 590 days and the peak-to-peak amplitudes are $\sim$\,3\,mag in G band, which means that they are genuine long-period large-amplitude variables. The variability information supports an AGB nature of these sources. In contrast to the case of IRAS~18319-1001, their detached positions in the G$-$RP HR diagram with respect to the bulk of the \G\ OH/IR sample are therefore not caused by incorrect identifications of their \G\ counterpart, but most probably by problems with the \G\ parallax and/or brightness determination.

In addition to IRAS~18319-1001, a group of six sources (\object{IRAS 18413-4616}, \object{MSB 38}, \object{IRAS 18006-3213}, \object{V840 Aql}, \object{V718 Cyg}, and \object{V CVn}) is detached from the main OH/IR star locus in the left panel of Fig.~\ref{fig:hrdiag} due to their bluer BP$-$RP colours (G$_{abs}$\,$\sim$\,--4.6 to --2.2\,mag, and BP$-$RP\,$\sim$\,2.3-3.4\,mag).
 \G\ G-band light curves are available for all these sources. Four of them (IRAS 18413-4616, MSB 38, V840 Aql, and IRAS 18006-3213) show an irregular variability with a low peak-to-peak amplitude ($<$0.3\,mag). V CVn is a semi-regular variable with a period of $\sim$\,200\,d \citep{Cadmus24} and a photometric peak-to-peak variability of 1 mag in the G band. The remaining two sources, V718 Cyg and IRC +00363, show Mira-like variability with long periods ($\sim$\,514\,d and $\la$\,950\,d, respectively) and a G-band photometric peak-to-peak variability of $\sim$1\,mag.

To summarise, the sources with the bluest colours in the HR diagrams are likely a mix of AGB stars with low-amplitude irregular and large-amplitude semi-regular and regular variability. A general study of the variability properties of the sources in the \G\ OH/IR sample will be presented in a forthcoming paper after the \G\ catalogue of Galactic AGB stars is complete.

\section{Conclusions}
\label{sec:conclusions}

We presented the first step towards the construction of the \G\ catalogue of Galactic AGB stars by providing a sample of OH/IR stars detected by \G\ DR3. We cross-matched the list of sources with detected OH maser emission (OH/IR stars) from the most recent version of the database of circumstellar OH masers of \citet{Engels24} with AllWISE, 2MASS, \G\ DR3, and other catalogues in order to retrieve astrometric and photometric information that we used to construct the SEDs of the objects from the near-ultraviolet to the far-infrared. We identified $\sim$1500 OH maser sources with \G\ detections, while more than 900 OH maser sources in the database with infrared identifications are too obscured to be detected by \G. Whenever distances based on trustworthy parallaxes were available, we obtained the interstellar extinction to each source and used it to deredden the observed SED. We then used the GRAMS model grid to fit the SEDs and obtain bolometric fluxes. Stellar luminosities were also determined for sources with reliable distances.

The AGB stars were separated from other OH maser sources based on  their SED shape, the fitting results, and information from the literature. The final `\G\ OH/IR star sample' contains 1172 sources, 222 of which have distance and luminosity estimations. These latter sources represent a bright and local subsample, given that most lie at distances within 2 kpc. As discussed by \cite{Lebzelter23}, a radius of 2 kpc qualifies as a reasonable border for the solar neighbourhood, in which most of the objects with good \G\ parallaxes are found.

For more than 80\% of the \G\ OH/IR stars, no parallaxes in the \G\ DR3 release match our adopted quality criteria because of circumstellar and interstellar extinction and, quite likely, because they are located beyond the solar neighbourhood. The OH maser database is dominated by a few OH maser surveys that were conducted at low Galactic latitudes and reached the Galactic centre at $\sim8$ kpc, which is reflected by the concentration of the \G\ OH/IR star sample towards the inner Galactic plane. In contrast, and as expected, the local subsample is evenly distributed on the sky.

The causes for the lack of good-quality \G\ parallaxes of the \G\ OH/IR stars is also evident from the  brightness and colour distributions. The local subsample is bright (mostly $G \la 12$), while in general, the \G\ OH/IR stars are optically faint ($(12 \la G \la 22)$. The G-band brightness distribution continuously increases up to G\,$\sim$\,19.5, and this would probably continue if the sizeable fraction of OH maser sources that is not detected by \G\ were not excluded by the survey detection limit. When circumstellar and interstellar extinction becomes significant, the SEDs become redder. While the local subsample has mostly colours $J-K_s \la 2$, the remaining \G\ OH/IR stars have typical colours $2 \la J-K_s \la 7$.

The \G\ OH/IR star sample is a well-curated sample of O-rich AGB stars that can be used as a benchmark sample for further studies of LPV with \G. It is the first sample in the \G\ Catalogue of Galactic AGB Stars we are assembling. In the next steps, a sample of carbon-rich AGB stars and another sample of oxygen-rich AGB stars without OH maser detections will be added.

\section*{Data availability}
\label{sec:data}

Tables \ref{tab:OHIR} and \ref{tab:rejected} are available in electronic form at the CDS via anonymous ftp to cdsarc.u-strasbg.fr (130.79.128.5) or via \url{http://cdsweb.u-strasbg.fr/cgi-bin/qcat?J/A+A/}. 

The Gaia Catalogue of Galactic OH/IR Stars can also be queried via an online service available at the Spanish Virtual Observatory (SVO) website (see Appendix \ref{sec:Append}).

\begin{acknowledgements}
Authors thank to John Fabio Aguilar Sánchez for the invaluable and useful discussion on distance uncertainties, and to Raquel Murillo-Ojeda for the help in the preparation of the SED graphs for the archive. 
This work is funded by ESA through the Faculty of the European Space Astronomy Centre (ESAC) - Funding reference 4000139151/22/ES/CM, and by grant No. PID2020-112949GB-I00 by the Spanish Ministry of Science, Innovation/State Agency of Research MCIN/AEI/10.13039/501100011033.

This publication has made use of data from the European Space Agency (ESA) mission \G\ (https://www.cosmos.esa.int/gaia), processed by the Gaia Data Processing and Analysis Consortium (DPAC, https://www.cosmos.esa.int/web/gaia/dpac/consortium); funding for the DPAC has been provided by national institutions, in particular the institutions participating in the Gaia Multilateral Agreement. 
It also made use of data products from the Wide-field Infrared Survey Explorer (WISE), which is a joint project of the University of California, Los Angeles, and the Jet Propulsion Laboratory/California Institute of Technology, funded by the National Aeronautics and Space Administration; and from the Two Micron All Sky Survey (2MASS), which is a joint project of the University of Massachusetts and the Infrared Processing and Analysis Center/California Institute of Technology, funded by the National Aeronautics and Space Administration and the National Science Foundation.

This work used as well data from the following catalogues and surveys: ESO's Visible and Infrared Survey Telescope for Astronomy (VISTA); the Digitized Sky Survey 2 (DSS-2), produced at the Space Telescope Science Institute under US Government grant NAG W-2166; the Galaxy Evolution Explorer GALEX-GR5 catalogues, a NASA mission led by the California Institute of Technology; the Deep Near-infrared southern Sky survey (DENIS); the AKARI/IRC and the AKARI/FIS mid-infrared surveys (ISAS/JAXA); the MSX6C Infrared Point Source Catalog; and the IRAS Point Source Catalog (NASA). We also made use of data and images from the Pan-STARRS1 Surveys (PS1), which have been made possible through contributions by the Institute for Astronomy, the University of Hawaii, the Pan-STARRS Project Office, the Max-Planck Society and its participating institutes, the Max Planck Institute for Astronomy, Heidelberg and the Max Planck Institute for Extraterrestrial Physics, Garching, The Johns Hopkins University, Durham University, the University of Edinburgh, the Queen's University Belfast, the Harvard-Smithsonian Center for Astrophysics, the Las Cumbres Observatory Global Telescope Network Incorporated, the National Central University of Taiwan, the Space Telescope Science Institute, the National Aeronautics and Space Administration under Grant No. NNX08AR22G issued through the Planetary Science Division of the NASA Science Mission Directorate, the National Science Foundation Grant No. AST-1238877, the University of Maryland, Eotvos Lorand University (ELTE), the Los Alamos National Laboratory, and the Gordon and Betty Moore Foundation. 

We used the following VO-compliant tools: Topcat \citep{Taylor05}; Aladin, developed at the Centre de Don\'ees astronomiques de Strasbourg (CDS), France; ESASky \citep{Baines2017}, developed at ESA/ESAC, Madrid, Spain, by the ESAC Science Data Centre (ESDC); and VOSA \citep{Bayo08}, developed under the Spanish Virtual Observatory (https://svo.cab.inta-csic.es) project funded by MCIN/AEI/10.13039/501100011033/ through grant PID2020-112949GB-I00. VOSA has been partially updated by using funding from the European Union's Horizon 2020 Research and Innovation Programme, under Grant Agreement nº 776403 (EXOPLANETS-A). This research also made use of the VizieR catalogue access tool and the SIMBAD database, both developed and maintained at the CDS, and of the ADS bibliographic services.
\end{acknowledgements}


\bibliographystyle{aa}
\bibliography{references}


\appendix

\onecolumn

\section{Online catalogue service}
\label{sec:Append}


\begin{table*}[h!]
  \caption{Description of the catalogue.}
  \label{tab:cat}
  \footnotesize
  \begin{center}
  \begin{tabular}{lll}
    \hline
    \hline
    \noalign{\smallskip}
Label & Unit & Description \\
    \noalign{\smallskip}    
\hline
    \noalign{\smallskip}                                           
RA\_(ICRS) & deg & \G-DR3 right ascension (Epoch J2016) \\ 
DEC\_(ICRS) & deg & \G-DR3 declination (Epoch J2016) \\ 
Gaia\_DR3\_ID & & {\it Gaia}-DR3 unique source identifier\\ 
SIMBAD\_ID& & SIMBAD source identifier \\ 
WISE\_ID& & WISE source identifier \\ 
2MASS\_ID && 2MASS source identifier \\ 
G & mag & \G-DR3 $G$ band mean magnitude \\    
BP & mag & \G-DR3 integrated $G_{BP}$ band mean magnitude \\    
RP & mag & \G-DR3 integrated $G_{RP}$ band mean magnitude \\    
parallax & mas & \G-DR3 parallax \\    
parallax\_error & mas & \G-DR3 standard error of parallax \\ 
distance$^{*}$ & pc & Geometric distance estimated by \cite{Bailer-Jones21} (see text) \\
A$_v$$^{*}$ & mag & Interstellar extinction estimated by \cite{Lallement22} (see text) \\
Fbol & erg/cm$^2$/s & Bolometric flux obtained from the SED fitting (see text) \\
Luminosity$^{*}$ & \Lsun & Stellar luminosity obtained from the bolometric flux and the distance\\
Sample$^{**}$ & & \G\ sample at which the source belongs (OH/IR, C-rich, O-rich)\\
SED\_phot & & URL link to a table containing the observational photometry of the source \\
SED\_graph & & URL link to the source SED graph \\
 \noalign{\smallskip}                                            
    \hline                                                          
\noalign{\smallskip}                                            
    \hline                                                          
 \end{tabular}                                                     
\end{center}  
Notes: 

$^{*}$For those objects with good parallax (see text)

$^{**}$The \G\ OH/IR star sample has been defined in this work. The \G\ C-rich and the \G\ O-rich samples are, respectively, a sample of C-rich AGB stars and O-rich AGB stars with no OH maser emission detected having \G\ counterparts that will be defined in forthcoming papers. 

\end{table*}

To help the astronomical community use the \G\ Catalogue of Galactic AGB Stars, we have developed an online catalogue service that can be accessed from a webpage\footnote{\footnotesize \url{http://svocats.cab.inta-csic.es/gaiaagb/}} or through a Virtual Observatory ConeSearch\footnote{\footnotesize e.g. \url{http://svocats.cab.inta-csic.es/gaiaagb/cs.php?RA=301.708&DEC=-67.482&SR=50&VERB=2}}. Table\,\ref{tab:cat} lists the data accessible through this service.

The service implements a simple search interface that permits queries by coordinates and radius as well as by other parameters of interest. The user can also select the maximum number of sources to return (with values from 10 to unlimited).

The result of the query is a HTML table with all the sources found in the catalogue fulfilling the search criteria, which can be downloaded as a VOTable or as a CSV file. Detailed information on the output fields can be obtained placing the mouse over the question mark (`?') located close to the name of the column. 
The results of a query can be easily transferred to other VO applications (e.g. Topcat) thanks to the Simple Application Messaging Protocol (SAMP)\footnote{\url{http://www.ivoa.net/documents/SAMP/}}, which allows Virtual Observatory applications to communicate with each other in a seamless and transparent manner.

The service has been designed as a live archive. Its initial version uses the data published in this paper. In following versions, it will be extended with more AGB star samples that will be published in forthcoming papers. Additional information, such as source variability properties obtained from \G\ DR3 light curves, will also be eventually added.

\end{document}